\documentclass{elsart5p}

\usepackage{epsfig}

\usepackage{amssymb}

\usepackage{xspace}

\def\cref#1{Chapt.\,\ref{#1}}
\def\Cref#1{Chapter~\ref{#1}}

\def\fref#1{Fig.\,\ref{#1}}

\def\1{\footnotemark[1]}
\def\and{\& }

\def\deg{$^\circ$\xspace}

\def\gcm2{g/cm$^2$\xspace}

\def\Section#1{\section{#1}}

\begin{document}

\begin{frontmatter}



\title{Investigation of the Properties of Galactic Cosmic Rays with the
       KASCADE-Grande Experiment} 


\author[2,16]{J.R. H\"orandel},
\author[1]{W.D. Apel},
\author[2,14]{J.C.~Arteaga},
\author[1]{F.~Badea},
\author[1]{K.~Bekk},
\author[5]{M.~Bertaina},
\author[1,2]{J.~Bl\"umer},
\author[1]{H.~Bozdog},
\author[7]{I.M.~Brancus},
\author[8]{M.~Br\"uggemann},
\author[8]{P.~Buchholz},
\author[5,9]{E.~Cantoni},
\author[5]{A.~Chiavassa},
\author[2]{F.~Cossavella},
\author[1]{K.~Daumiller},
\author[2,15]{V.~de Souza},
\author[5]{F.~Di~Pierro},
\author[1]{P.~Doll},
\author[1]{R.~Engel},
\author[1]{J.~Engler},
\author[1]{M.~Finger},
\author[11]{D.~Fuhrmann},
\author[9]{P.L.~Ghia},
\author[1]{H.J.~Gils},
\author[11]{R.~Glasstetter},
\author[8]{C.~Grupen},
\author[1]{A.~Haungs},
\author[1]{D.~Heck},
\author[1]{T.~Huege},
\author[1]{P.G.~Isar},
\author[11]{K.-H.~Kampert},
\author[2]{D.~Kang},
\author[8]{D.~Kickelbick},
\author[1]{H.O.~Klages},
\author[12]{P.~{\L}uczak},
\author[1]{H.J.~Mathes},
\author[1]{H.J.~Mayer},
\author[7]{B.~Mitrica},
\author[9]{C.~Morello},
\author[5]{G.~Navarra},
\author[1]{S.~Nehls},
\author[1]{J.~Oehlschl\"ager},
\author[1]{S.~Ostapchenko},
\author[8]{S.~Over},
\author[7]{M.~Petcu},
\author[1]{T.~Pierog},
\author[1]{H.~Rebel},
\author[1]{M.~Roth},
\author[1]{H.~Schieler},
\author[1]{F.~Schr\"oder},
\author[13]{O.~Sima},
\author[2]{M.~St\"umpert},
\author[7]{G.~Toma},
\author[9]{G.C.~Trinchero},
\author[1]{H.~Ulrich},
\author[1]{A.~Weindl},
\author[1]{J.~Wochele},
\author[1]{M.~Wommer},
\author[12]{J.~Zabierowski}

\address[2]{Institut f\"ur Experimentelle Kernphysik,
 Universit\"at Karlsruhe, Germany}
\address[1]{Institut f\"ur Kernphysik, Forschungszentrum Karlsruhe, Germany}
\address[5]{Dipartimento di Fisica Generale dell'Universit{\`a} di Torino, Italy}
\address[7]{National Institute of Physics and Nuclear
 Engineering, Bucharest, Romania}
\address[8]{Fachbereich Physik, Universit\"at Siegen,
 Germany}
\address[9]{Istituto di Fisica dello Spazio Interplan
etario, INAF Torino, Italy}
\address[11]{Fachbereich Physik, Universit\"at Wuppertal, Germany}
\address[12]{Soltan Institute for Nuclear Studies, Lodz, Poland}
\address[13]{Department of Physics, University of Bucharest, Bucharest, Romania}
\address[16]{\small now at: Radboud University Nijmegen, Department of
        Astrophysics, P.O. Box 9010, 6500 GL Nijmegen, The Netherlands }
\address[14]{\small now at: Universidad Michoacana, Morelia, Mexico}
\address[15]{\small now at: Universidade de S\~ao Paulo, Instituto de
             F\'{\i}sica de S\~ao Carlos, Brasil}

\begin{abstract}
The properties of galactic cosmic rays are investigated with the KASCADE-Grande
experiment in the energy range between $10^{14}$ and $10^{18}$~eV.
Recent results are discussed. They concern mainly the all-particle energy
spectrum and the elemental composition of cosmic rays.
\end{abstract}

\journal{Nuclear Instruments and Methods A}
\end{frontmatter}

\Section{Introduction}
To reveal the origin of galactic cosmic rays is the main objective of the
KASCADE and KASCADE-Grande experiments.  The results of KASCADE contributed
significantly to the understanding of the origin of the knee in the
all-particle energy spectrum of cosmic rays at energies around
$4\cdot10^{15}$~eV.  It could be shown that the knee is caused by a fall-off in
the flux of light nuclei \cite{ulrichapp,Apel:2008cd}.  Energy spectra for five
dominant elemental groups could be reconstructed (p, He, CNO, Si, and Fe), they
exhibit a fall-off behavior approximately proportional to the nuclear charge of
the elemental groups.  In the energy region between $10^{17}$ and $10^{18}$~eV
a transition from a galactic to an extra-galactic origin of cosmic rays is
expected \cite{behreview}.  Main focus of KASCADE-Grande is to understand the
end of the galactic cosmic-ray component through detailed investigations of the
energy spectrum and the mass composition in this energy region.

KASCADE-Grande comprises 37 detector stations equipped with plastic
scintillators, covering an area of 0.5~km$^2$ to measure the electromagnetic
shower component \cite{grande}.
It also includes the original KASCADE experiment, consisting 
of several detector systems \cite{kascadenim}. A $200 \times
200$~m$^2$ array of 252 detector stations, equipped with scintillation
counters, measures the electromagnetic and, below a lead/iron shielding, the
muonic parts of air showers.  
A 130~m$^2$ streamer tube detector, shielded by a soil-iron absorber serves to
reconstruct the tracks of high-energy ($E_\mu>0.8$~GeV) muons \cite{mtdnim}.
An iron sampling calorimeter of $16 \times 20$~m$^2$ area detects hadronic
particles \cite{kalonim}.  It has been calibrated with a test beam at the SPS
at CERN up to 350~GeV particle energy \cite{kalocern}.

\begin{figure}\centering
 \epsfig{file=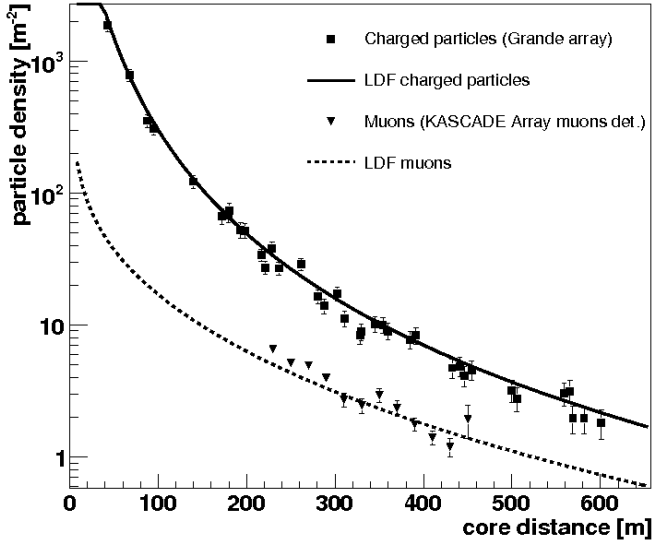, width=\columnwidth}
 \caption{Measured lateral distribution of a single event for charged particles
          and muons \cite{icrc09-dipierro}.}
 \label{lat}
\end{figure}

As an example for a measured air shower, the lateral distributions for charged
particles and for muons are shown in \fref{lat} \cite{icrc09-dipierro}. The
charged-particle density is measured with the Grande detectors and the muonic
component by the shielded array detectors of KASCADE.

\Section{Test of hadronic interaction models}

The astrophysical interpretation of air shower data requires detailed
knowledge of the properties of hadronic interactions at energies and
kinematical ranges beyond the capabilities of present-day accelerator
experiments. Therefore, air shower data are used to constrain hadronic
interaction models used in air shower simulations, such as the CORSIKA
\cite{corsika} code.  Several hadronic interaction models have been
systematically tested over the last decade.  

First quantitative tests \cite{Horandel:1998br,wwicrc99,wwtestjpg} established
QGSJET~98 \cite{qgsjet} as the most compatible code. Similar conclusions have
been drawn for the successor code QGSJET~01 \cite{jensjpg}.  Recently, a new
method has been developed to measure the attenuation of hadrons in air showers
\cite{atteas}. This method is very sensitive to inelastic hadronic cross
sections. The investigations indicate, that the inelastic cross sections in
QGSJET~01 are slightly too large ($\approx5\%$ at $10^6$~GeV).  Predictions of
QGSJET II \cite{qgsjet2,qgsII,qgsjetII} exhibit problems when compared to air
shower data \cite{icrc09-hoerandel}, in particular, the predicted
electron-hadron correlations are not compatible with the measurements.

Predictions of SIBYLL~1.6 \cite{sibyll16} were not compatible with air shower
data, in particular there were strong inconsistencies for hadron-muon
correlations. These findings stimulated the development of SIBYLL~2.1
\cite{sibyll21}. This model proved to be very successful, the predictions of
this code are fully compatible with KASCADE air shower data
\cite{jenskrakow,jenspune,jensjpg}. 

Investigations of the VENUS \cite{venus} model revealed some inconsistencies in
hadron-electron correlations \cite{wwtestjpg}.  The predictions of {\sc
neXus~2} \cite{nexus} were found to be incompatible with the KASCADE data, in
particular, when hadron-electron correlations have been investigated
\cite{jensjpg}.
Recently, predictions of the interaction model EPOS~1.61
\cite{epos,eposmerida,epos2} have been compared to KASCADE air shower data
\cite{epostest}.  The analysis indicates that EPOS~1.61 delivers not enough
hadronic energy to the observation level and the energy per hadron seems to be
too small.  These findings stimulated the development of a new version EPOS~1.9
\cite{icrc09-pierog}.  Corresponding investigations with this new version are
under way and results are expected to be published soon.

Analyses of the predictions of the DPMJET model yield significant problems in
particular for hadron-muon correlations for the version DPMJET~2.5
\cite{dpmjet}, while the newer version DPMJET~2.55 is found to be compatible
with air shower data \cite{jensjpg}.

Presently, the most compatible predictions are obtained from the models
QGSJET~01 and SIBYLL~2.1.

\Section{Muons in air showers}

The lateral distribution of muons ($E_\mu>230$~MeV) is investigated using the
shielded detectors of the original KASCADE array up to distances to the shower
axis exceeding 600~m \cite{icrc09-fuhrmann}. A function suggested by Lagutin
and Raikin is used to described the measured muon densities as a function of
the distance to the shower axis. This method allows to reconstruct the total
number of muons in air showers with an accuracy better than 20\%.

A comparison of measured muon densities to predictions of air shower
simulations indicates that these observable is sensitive to differences in
hadronic interaction models \cite{icrc09-desouza}. The muon density has been
measured as a function of the total number of electrons in the showers, i.e.\
as a function of primary energy.

\begin{figure}\centering
 \epsfig{file=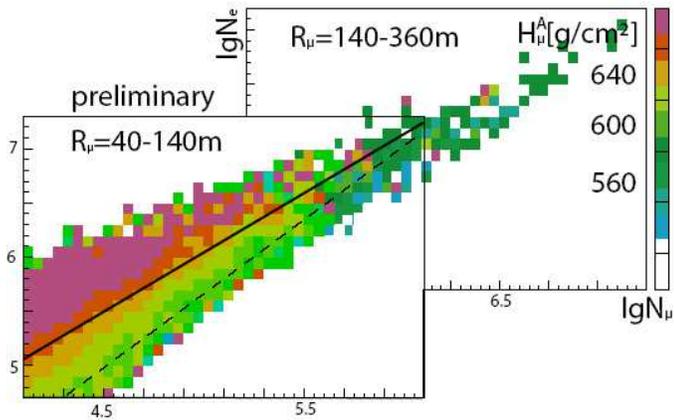, width=\columnwidth}
 \caption{Effective depth of the muon production $H_\mu^A$ in the $N_e-N_\mu$
	  plane for showers with zenith angle smaller than 18\deg
	  \cite{icrc09-doll}.  Diagrams are overlayed for separate KASCADE and
          KASCADE-Grande analyses.}
 \label{mph}
\end{figure}

Muons with energies exceeding 800~MeV are investigated with the muon tracking
detector. Lateral distributions have been measured, they can be described by a
Lagutin-like function \cite{icrc09-luczak}.
The information of the reconstructed muon tracks is used to obtain radial and
tangential angles with respect to the shower axis. These quantities, in turn,
are applied to infer the pseudorapidity distributions of secondary particles
in hadronic interactions which decay into muons \cite{icrc09-zabierowski}.

The muon tracks are also used to measure the production height of muons in the
atmosphere by means of triangulation \cite{icrc09-doll}. The corresponding
production depth of muons in the atmosphere is depicted in \fref{mph} in the
$N_e-N_\mu$ plane.  The figure combines two separate analyses for the energy
ranges of KASCADE and KASCADE-Grande, respectively.  In the electron-muon
number plane light particles are expected to be to the upper left of the main
diagonal. In this region large muon production depths are reconstructed. This
indicates light particles, penetrating deep into the atmosphere.  The measured
air showers are classified in four intervals of the muon production depth. For
these intervals the primary energy spectrum is reconstructed, based on the
measured number of electrons and muons.  This yields energy spectra for four
elemental groups. The spectra exhibit a fall-off for the light component at low
energies (several $10^{15}$~eV), while the all-particle spectrum at high
energies is dominated by heavy nuclei \cite{icrc09-doll}.

\Section{All-particle energy spectrum}

Different methods are applied to the data to reconstruct the all-particle
energy spectrum.

The density of charged particles at large distances from the shower axis is a
good mass-independent estimator for the shower energy \cite{icrc09-toma}. The
particular optimum radial distance depends on the layout and altitude of the
detector array, for KASCADE-Grande a distance of 500~m yields best values and
we use the charged-particle density $S(500)$ at 500~m from the shower axis as
energy estimator. The constant intensity cut method is applied to evaluate the
attenuation behavior of the $S(500)$ values. Thus, the all-particle energy
spectrum is reconstructed.

Another analysis takes advantage of the total number of charged particles
measured for each air shower \cite{icrc09-kang}. The shower size spectra
measured in different zenith angle intervals are used to obtain the attenuation
length for this observable. The constant intensity cut method is again applied
to correct for attenuation effects and the primary energy spectrum is obtained.

In a similar way the muon data have been analyzed \cite{icrc09-arteaga}. The
constant intensity cut method is used to correct for the attenuation of the
measured total number of muons in different zenith angle intervals.  Again,
predictions of air shower simulations are used to establish an energy scale for
the observed quantities and the all-particle energy spectrum is derived from
the observed number of muons.

The last analysis discussed here, combines information from charged particles
and muons \cite{icrc09-bertaina}. Independent fits to the lateral distributions
of charged particles and muons allow to extract the number of charged particles
and number of muons for each shower.  These quantities are used to assign an
energy to each shower on an event-by-event basis. Thus, the all-particle energy
spectrum is inferred.

\begin{figure*}\centering
 \epsfig{file=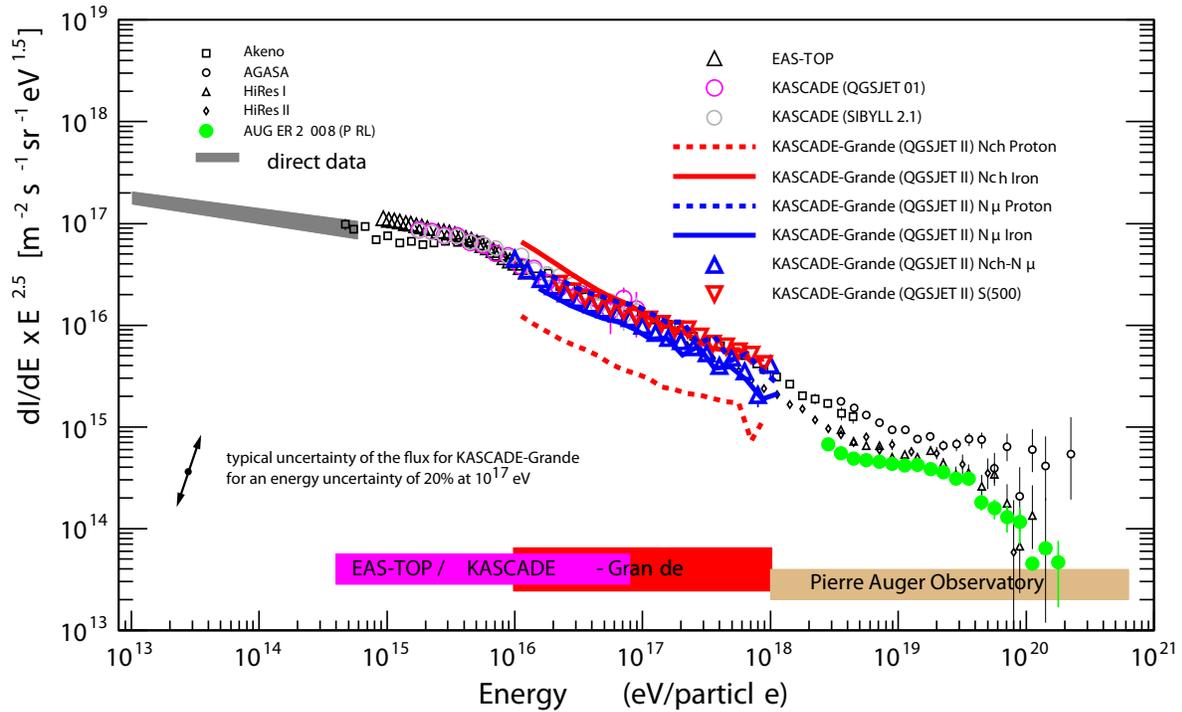, width=0.85\textwidth}
 \caption{All-particle energy spectrum of cosmic rays. Spectra obtained with
          different methods applied to KASCADE-Grande data are compared to
          results of other experiments  \cite{kascadeicrc09}.}
 \label{kgall}
\end{figure*}

The results obtained with the different methods are compiled in \fref{kgall}.
The reconstructed all-particle flux is shown as a function of the primary
energy.  The arrow indicates a typical systematic uncertainty of the energy
scale of the order of 20\%. Since the elemental composition is presently not
clear in this energy region, for some methods two spectra are plotted, one for
proton induced showers and a second one for iron induced showers.  The
different methods agree well for a heavy dominated mass composition around
energies of $10^{17}$~eV \cite{icrc09-haungs}.  The spectra obtained in the
energy range between $10^{16}$ and $10^{18}$~eV smoothly extend the energy
spectrum as measured by KASCADE to high energies.  The new results match as
well the flux obtained by other experiments, shown in the figure as well.

\Section{Composition}

First analyses have started to reveal the elemental composition at the end of
the galactic component \cite{icrc09-cantoni}.  They use the measured number of
electrons and muons and indicate that these observables are indeed sensitive to
the elemental composition. At present, the measured data can be described by a
fit of three elemental groups. A comparison of the measured electron-to-muon
number ratio from KASCADE-Grande to similar data obtained with KASCADE yields
consistent results.

Ultimate goal is a deconvolution of the measured size spectra for the
electromagnetic/charged and the muonic shower components, similar to the
unfolding procedure applied to KASCADE data \cite{ulrichapp} to obtain energy
spectra for separate elemental groups at energies up to $10^{18}$~eV.

\Section{Conclusions and outlook}
The recent results from the KASCADE-Grande experiment indicate that we make
substantial progress in measuring the energy spectrum and the elemental
composition of cosmic rays in the energy region between $10^{16}$ and
$10^{18}$~eV. The new data will significantly contribute to the understanding
of the end of the galactic component and the transition to an extra-galactic
component. The data will be crucial to distinguish between different
astrophysical scenarios.


\end{document}